\documentclass[twocolumn]{aastex701}
\usepackage[utf8]{inputenc}
\usepackage{amsmath}
\usepackage{natbib}
\usepackage{epstopdf}
\usepackage{gensymb}
\usepackage{CJKutf8}
\usepackage{rotating}
\usepackage{graphicx}

\shorttitle{PAH-Metallicity Relation II: the Smallest and Stubbornest PAHs}
\shortauthors{Whitcomb et al.}

\newcommand{\um}{\,\micron}
\newcommand{\hii}{\textsc{Hii}}

\newcommand{\kjysr}{kJy\,sr$^{-1}$}
\newcommand{\mjysr}{MJy\,sr$^{-1}$}
\newcommand{\jdu}{10$^{-7}$\,W\,m$^{-2}$\,sr$^{-1}$}

\newcommand{\totpah}{$\Sigma$PAH}
\newcommand{\pahtir}{$\Sigma$PAH/TIR}
\newcommand{\qpah}{$q_{\rm {PAH}}$}

\newcommand{\pahthree}{PAH\,3.3/$\Sigma$PAH}

\newcommand{\carbons}{$N_{\mathrm{C}}$}
\newcommand{\aone}{\ensuremath{a_{\rm 01}}}
\newcommand{\amin}{$a_{\rm min}$}

\newcommand{\twelog}{12~+~[O/H]}

\newcommand{\zsolar}{$Z_{\odot}$}

\newcommand{\zthresh}{$Z_\mathrm{th}$}

\newcommand{\fpah}{F335M$_{\rm PAH}$}
\newcommand{\threethree}{3.3\,\micron}

\newcommand{\spitzer}{Spitzer}
\begin{document}

\title{The Metallicity Dependence of PAH Emission in Galaxies II: \\Insights from JWST/NIRCam Imaging of the Smallest Dust Grains in M101}

\author[0000-0003-2093-4452]{Cory M. Whitcomb}
\affil{Ritter Astrophysical Research Center, Department of Physics \& Astronomy, University of Toledo, Toledo, OH 43606, USA}
\email[show]{coryw777@gmail.com}

\author[0000-0003-1545-5078]{J.-D. T. Smith}
\affil{Ritter Astrophysical Research Center, Department of Physics \& Astronomy, University of Toledo, Toledo, OH 43606, USA}
\email{JD.Smith@utoledo.edu}

\author[0000-0003-1356-1096]{Elizabeth Tarantino}
\affiliation{Space Telescope Science Institute, 3700 San Martin Drive, Baltimore, MD 21218, USA}
\email{etarantino@stsci.edu}

\author[0000-0002-4378-8534]{Karin Sandstrom}
\affil{Department of Astronomy \& Astrophysics, University of California, San Diego,\\ 9500 Gilman Drive, La Jolla, CA 92093, USA}
\email{karin.sandstrom@gmail.com}

\author[0000-0001-8490-6632]{Thomas S.-Y. Lai\begin{CJK*}{UTF8}{bsmi} (賴劭愉)\end{CJK*}}
\affil{IPAC, California Institute of Technology, 1200 E. California Boulevard, Pasadena, CA 91125, USA}
\email{shaoyu@ipac.caltech.edu}

\author[0000-0003-3498-2973]{Lee Armus}
\affil{IPAC, California Institute of Technology, 1200 E. California Boulevard, Pasadena, CA 91125, USA}
\email{lee@ipac.caltech.edu}

\author[0000-0002-5480-5686]{Alberto Bolatto}
\affil{Department of Astronomy, University of Maryland, College Park, MD 20742, USA}
\email{bolatto@umd.edu}

\author[0000-0003-4850-9589]{Martha Boyer}
\affiliation{Space Telescope Science Institute, 3700 San Martin Drive, Baltimore, MD 21218, USA}
\email{mboyer@stsci.edu}

\author[0000-0002-5782-9093]{Daniel~A.~Dale}
\affiliation{Department of Physics and Astronomy, University of Wyoming, Laramie, WY 82071, USA}
\email{DDale@uwyo.edu}

\author[0000-0002-0846-936X]{Bruce T. Draine}
\affil{Department of Astrophysical Sciences, Princeton University, Princeton, NJ 08544, USA}
\email{draine@astro.princeton.edu}

\author[0000-0001-7449-4638]{Brandon S. Hensley}
\affil{Jet Propulsion Laboratory, California Institute of Technology, 4800 Oak Grove Drive, Pasadena, CA, USA}
\email{brandon.s.hensley@jpl.nasa.gov}

\author[0000-0002-7064-4309]{Desika Narayanan}
\affil{Department of Astronomy, University of Florida, 211 Bryant Space Sciences Center, Gainesville, FL 32611, USA}
\email{desika.narayanan@ufl.edu}

\author[0000-0001-6326-7069]{Julia Roman-Duval}
\affiliation{Space Telescope Science Institute, 3700 San Martin Drive, Baltimore, MD 21218, USA}
\email{duval@stsci.edu}

\author[0000-0003-0605-8732]{Evan D. Skillman}
\affiliation{University of Minnesota, Minnesota Institute for Astrophysics, School of Physics and Astronomy,\\ 116 Church Street, S.E., Minneapolis, MN 55455, USA}
\email{skill001@umn.edu}

\correspondingauthor{Cory M. Whitcomb}

\begin{abstract}

We explore the physical origins of the observed deficit of polycyclic aromatic hydrocarbons (PAHs) at sub-solar metallicity using JWST/NIRCam imaging of the nearby galaxy M101, covering regions from solar metallicity (\zsolar) down to 0.4\,\zsolar.  These maps are used to trace the radial evolution of the shortest-wavelength PAH feature at \threethree, which is emitted preferentially by the smallest PAHs ($<$100 carbon atoms). The fractional contribution of PAH \threethree\ to the total PAH luminosity (\totpah) increases by $3\times$ as metallicity declines, rising from $\sim$1\% to $\sim$3\% over the observed range, consistent with prior predictions from the inhibited grain growth model based on \spitzer\ spectroscopy.
We explore model refinements including photon effects and alternative size evolution prescriptions, and find that a modest amount of small grain photo-destruction remains possible, provided the grain size cutoff does not exceed $\sim$55 carbon atoms. 
The best-fit models predict \threethree$/$\totpah\ will rise to $\sim$$5.6-7.7\%$ at 10\%\,\zsolar. Surprisingly, even as \totpah\ drops significantly relative to the total infrared luminosity (TIR) as metallicity declines, \threethree/TIR alone \emph{rises}, potentially indicating the mass fraction of the smallest PAH grains increases as the total dust content in galaxies drops. The current model cannot fully reproduce this trend even if the unusually strong effects of changing radiation field hardness on \threethree/TIR are included. This may be evidence that the smallest PAHs are uniquely robust against destruction and inhibited growth effects. These results highlight the pivotal role that short-wavelength PAH emission can play in studies of low-metallicity and high-redshift galaxies.

\end{abstract}


\section{Introduction}\label{sec:intro}
The smallest carbonaceous dust grains that can survive in the interstellar medium (ISM) emit a strong vibrational feature at \threethree. These dust grains are commonly believed to be polycyclic aromatic hydrocarbons (PAHs) composed of about 20 -- 100 carbon atoms \citep{tiel08, D21}. The general population of PAHs extends to molecules composed of $\sim1,000$ carbon atoms which are theorized to be responsible for the strong, broad emission features in the 3 -- 18\um\ spectral range in observations of the ISM of galaxies. The emission from the \threethree\ PAH feature constitutes about 1 -- 3\% of all PAH emission power, and about 0.1\% of the total infrared luminosity of star-forming galaxies \citep{lai20}.

Several studies have found a significant deficit in the luminosity from PAH vibrational features relative to the total dust luminosity at sub-solar metallicity \citep[i.e., the PAH-metallicity relation or PZR, see][and references therein]{whitcomb24}. Many of these studies attribute the deficit to a decrease in the fractional abundance of PAHs relative to larger dust grains due to enhanced photo-destruction of smaller PAHs in the harder radiation fields of low-metallicity environments where dust shielding and ISM density are significantly lower \citep{madden06, gordon08}.
Recent studies have reported MIR emission from PAHs \citep{spilker23, witstok23} and UV attenuation by PAHs \citep{markov25, ormerod25} in high redshift galaxies.
It is crucial that we understand the local relationship between PAHs and metallicity to inform our interpretation of unexpectedly rapid dust formation at these early epochs.

The first Paper~In this series \citep[][Paper~I]{whitcomb24} used \spitzer/Infrared Spectrograph (IRS) spectroscopic maps to study the observed PZR for all PAH bands between 5 and 20\um\ in three nearby spiral galaxies: M101, NGC~628, and NGC~2403. We found that, when corrected for the effects of softer radiation fields in the galaxy's bulge, the ratio of total PAH luminosity (\totpah) to total infrared luminosity (TIR) is approximately constant at high metallicity, as seen in previous studies \citep{sutter24}. \pahtir\ remains constant until a threshold is reached at \zthresh~$\equiv$~0.63$~\pm~$0.03 of solar metallicity (\zsolar), below which \totpah/TIR decreases rapidly. By studying the evolution of PAH feature strengths as metallicity decreases, we found a general trend where the total PAH luminosity shifts from longer to shorter wavelength features.

To test potential explanations for the observed band-resolved PZR, we built on the latest dust grain models from \citet[hereafter D21]{D21} that depend on grain size and the illuminating radiation field. The leading model from Paper~I (W24 model) shifts average PAH sizes smaller and decreases the overall number of PAH grains as metallicity decreases. This behavior can be interpreted as a result of metallicity-dependent inhibited grain growth. In this scenario, PAH grains are theorized to grow through the accretion of carbon atoms and other PAH precursors in the dense ISM \citep{zhang25}, but this process is increasingly inhibited as the gas-phase metallicity decreases \citep{zhukovska13}. The W24 model predicts that the shortest-wavelength major PAH feature at \threethree, which was inaccessible with \spitzer\ spectroscopy, would vary most strongly with metallicity, increasing in fractional strength relative to all PAH power (\pahthree) by more than a factor of three between \zsolar\ and 0.4\,\zsolar.

In this paper, the second in the series, we combine new JWST/NIRCam imaging 
%(20 kpc x 4.5 kpc)
of the \threethree\ PAH feature with the \spitzer-IRS spectroscopy from Paper~I
%(20 kpc x 1.5 kpc)
in a deep radial strip across the nearby star-forming galaxy M101, covering metallicities from \zsolar\ to 0.4~\zsolar. With these data, we can study co-spatial variations in \emph{all} major PAH features as a function of metallicity at physical scales of $\sim350$\,pc. Including the \threethree\ feature allows us to trace the abundance of the smallest PAH grains in galaxies and place new constraints on the inhibited growth PZR model. We find that, as predicted in Paper~I, PAH \threethree/$\Sigma$PAH increases substantially as $Z/Z_\odot$ varies from 1 to 0.4.

\begin{sidewaysfigure*}
\includegraphics[width=\textwidth]{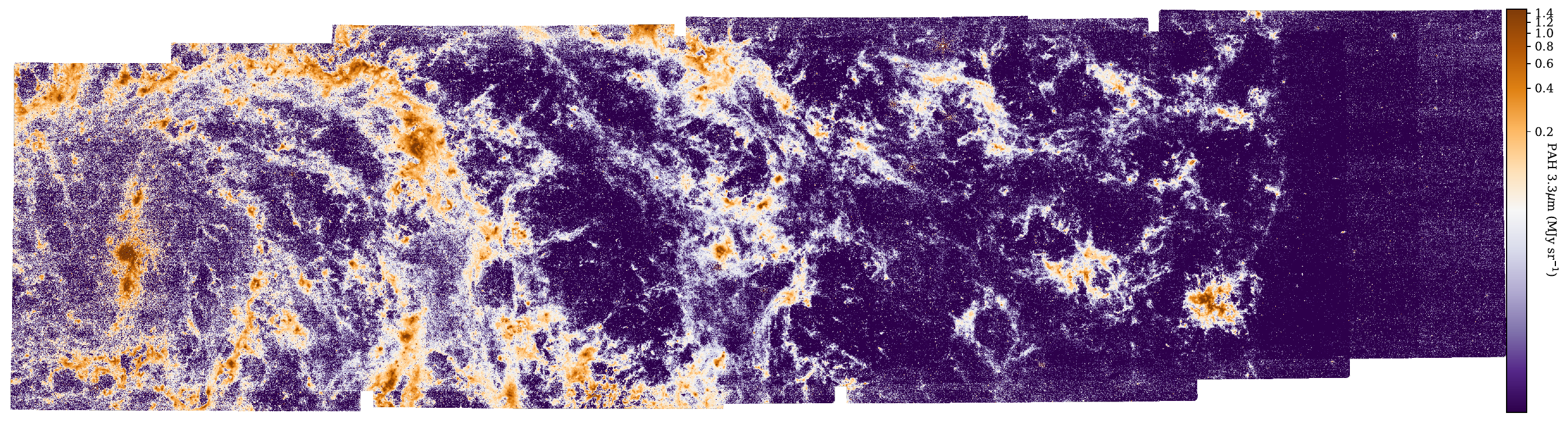}
\includegraphics[width=0.961\textwidth]{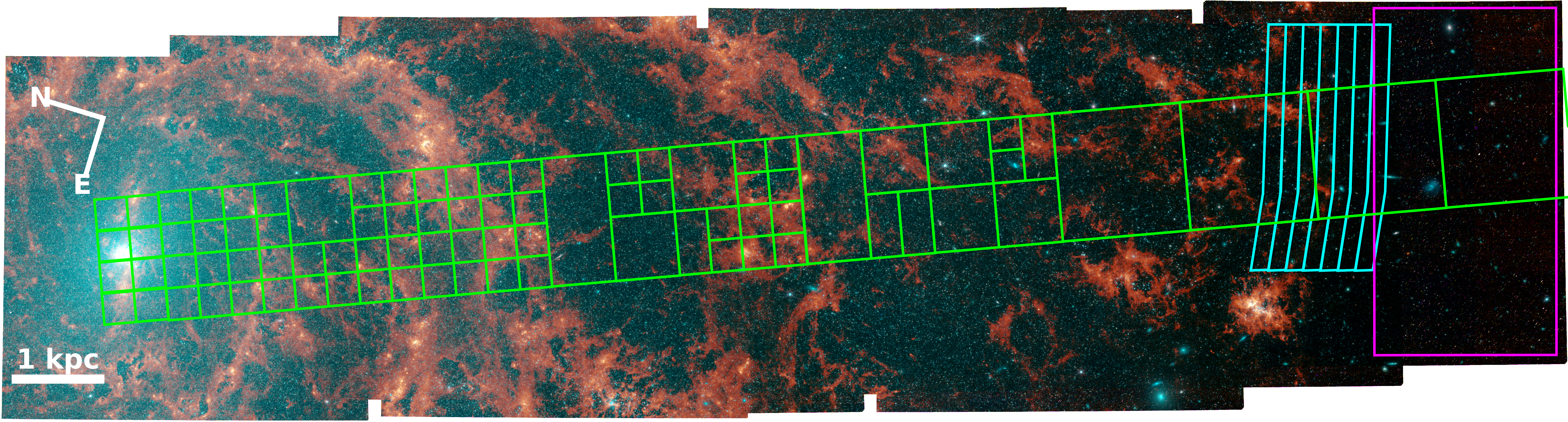}
\caption{\textbf{Top:} PAH \threethree\ emission isolated from F335M with the method adopted in this work (see \S\,\ref{sec:pahextract}). Note the apparent `cliff’ at the right of the image where galaxy structure seems to end abruptly (see \textbf{Appendix}\,\ref{sec:cliff}). \textbf{Bottom:} three-color image with PAH \threethree\ in red, F360M in green, and F300M in blue. Apertures defined in Paper~I are indicated in green, the apertures used to study the cliff are shown in cyan, and the off-galaxy aperture discussed in \S\,\ref{sec:anchor} is shown in magenta.
}
\label{figure:prettypics}
\end{sidewaysfigure*}

This paper is organized as follows. In \S\,\ref{sec:data} we describe our JWST/NIRCam data reduction and analysis process. In \S\,\ref{sec:pahextract} we describe the methods used to isolate PAH \threethree\ emission in the JWST/NIRCam data. In \S\,\ref{sec:obsresults} we describe the details of the observed trend between \pahthree\ and metallicity. In \S\,\ref{sec:dvresults} we describe our models of the observed trends in the JWST/NIRCam and \spitzer/IRS data. In \S\,\ref{sec:disc} we discuss the physical implications of our results, compare them with previous studies, and make predictions for future low-metallicity observations. In \S\,\ref{sec:concl} we outline our most significant conclusions.

\vspace{-7 pt}

\section{Data and Methods}\label{sec:data}

The primary data analyzed in this work are JWST/NIRCam \citep{Gardner+2023,Rieke+2023,Rigby+2023} photometric mosaics of M101 observed in Cycle 1 program 2452 (PI: J.D. Smith). This program was designed to overlap with the footprint of existing \spitzer\ spectroscopic maps presented in Paper~I. The region was observed in the F200W, F212N, F300M, F335M, and F360M filters. 
%Each medium- and wide-band filter was exposed for 47.25 minutes in total and 94.5 minutes in total with the narrow-band F212N filter.
Each filter was observed at 6 dither positions with 3 primary and 2 sub-pixel dithers in the 3TIGHT pattern \citep{jwstdithers}. There were 5 groups per integration for 6 integrations, readout using the SHALLOW2 pattern \citep{jwstreadout}.

\subsection{NIRCam Data Reduction}\label{sec:pipeline}

All data from program 2452 were retrieved from the Mikulski Archive for Space Telescopes (MAST) and reduced using version 1.17.1 of the JWST/NIRCam pipeline \citep{pipeline} with CRDS context jwst\textunderscore1322.pmap. Default parameters were used, and a 1/$f$ noise correction was applied to all Stage 2 images before mosaicing \footnote{\url{https://github.com/chriswillott/jwst/blob/master/image1overf.py}}.

We found two general alignment issues in the resulting mosaics: internal alignment between tiles within each mosaic was off by about 2\arcsec\ ($\sim55$ pixels in F200W), and there were not enough foreground stars in the Gaia DR3 catalog \citep{gaia, gaiadr3} at the outer edge of M101 to achieve sufficient absolute alignment between filters. We found that the internal alignment issue was a result of the guide star position used for some of the tiles. A total of four guide stars were used for each mosaic. The exposures that used one of the guide stars, N484003064, all exhibited an approximately 2\arcsec\ shift with respect to the tiles that used the other guide stars, which composes roughly one-third of the total tiles for each mosaic.

For the internal alignment issue, we used \texttt{jhat} \citep{jhat} to input an initial shift for all exposures that used N484003064 as a guide star. For the short-wavelength (SW) filters (F200W and F212N) we use the shifts (x, y) = (55 pix, -30 pix) and for the long-wavelength (LW) filters (F300M, F335M, and F360M) we use the shifts (x, y) = (25 pix, -16 pix). Once the exposures were pre-shifted, \texttt{jhat} alignment succeeded for all mosaics. The final alignment between each mosaic (SW and LW) is accurate to within about half a pixel. The absolute alignment of the tiles from the F300M filter was sufficient for our science goals (comparing with \spitzer/IRS spectroscopy) using the Gaia DR3 catalog stars at the center of M101. We aligned this mosaic to Gaia with \texttt{jhat} and generated a catalog from F300M to which the other images were matched.

\subsection{Ancillary Data}\label{sec:otherdata}

To place \threethree\ in context, we also used the \spitzer/IRS spectroscopic maps of M101 presented in Paper~I. The coverage of these maps is indicated by the green apertures shown in the bottom panel of Figure~\ref{figure:prettypics}. We used the same metallicity gradient for M101 as Paper~I which is derived from auroral-line measurements of \hii\ regions throughout the galaxy \citep{chaosVI}. We defined solar metallicity as \zsolar~$\equiv$~\twelog~=~8.69 \citep{asplund09} and we assumed that O/H is a good tracer of the abundance of all metal species and that the metallicity gradient is purely radial (see Paper~I Section 3.5 for further discussion of these assumptions). The metallicity for each region is given by: \twelog~$=(8.70~\pm~0.04)-(0.17~\pm~0.02)\times(R/R_{\rm e})$, where the metallicity decreases with a constant slope as a function of deprojected galactocentric radius ($R$) relative to the effective radius of the galaxy ($R_{\rm e}$ = 197.6\arcsec). With this gradient, our NIRCam footprint spans metallicities ranging from 1.02~$\pm$~0.02\,\zsolar\ to 0.39~$\pm$~0.01\,\zsolar. It is, however, important to note that these metallicities derived from auroral-line measurements may differ from strong-line metallicity calibrations which can vary up to 0.7~dex in absolute abundance \citep{chaosIV}.

In order to derive the TIR surface brightness in each of our apertures from Paper~I, we used photometric data from 24\,\micron\ through 160\,\micron. The 24\,\micron\ data is from the Multiband Imaging Photometer for \spitzer\ (MIPS), and the 70, 100, and 160\,\micron\ data from the Photodetector Array Camera and Spectrometer (PACS) of the Herschel Space Observatory were retrieved from the Key Insights on Nearby Galaxies: a Far-Infrared Survey with Herschel data release \citep[KINGFISH,][]{KINGFISH}. We then used the empirical calibration from \citet{galam13} -- with the equation and constants from their Table 3 -- to derive TIR surface brightness from a linear combination of MIPS~24\,\micron, and PACS 70, 100, and 160\,\micron\ photometry.

We additionally utilized radial profiles in the following images and maps, in ascending wavelength order: Galaxy Evolution Explorer (GALEX) far-ultraviolet (FUV 1500--1800\AA), Infrared Array Camera (IRAC) 3.6\um\ and 8\um, MIPS 24\um, and \textsc{Hi} 21\,cm. The FUV image used was from Guest Investigator Cycle 3 with ID
GI3\textunderscore 050008\textunderscore NGC5457 obtained from MAST, the IRAC and MIPS images were observed as part of the \spitzer\ Local Volume Legacy \citep[LVL,][]{dale09} and obtained from the \spitzer\ Heritage Archive, and the \textsc{Hi} moment zero maps were obtained from the \textsc{Hi} Nearby Galaxy Survey (THINGS) data release \citep{things}. The radial profiles were constructed using seven apertures (cyan in Figure~\ref{figure:prettypics}) where the middle aperture was drawn to fit the shape of the PAH~\threethree\ `cliff' (see \S\,\ref{sec:cliff}) with the width set by the limiting point-spread function (PSF) (MIPS~24\,\micron, FWHM~$\sim6\farcs4$). This aperture was dithered three times inward and three times outward to trace the trends in emission before and after the cliff.

% \begin{figure}
% \includegraphics[width=0.95\linewidth, height=0.9\textheight]{anchor_corrs_myrun.pdf}
% \caption{Correlations between IRAC~3.6\um\ and F300M (top), F335M (center), and F360M (bottom) for annular apertures described in \S\,\ref{sec:anchor}. The dashed gray lines indicate a linear fit to the data points dimmer than 25~\kjysr\ and the identity line is shown in black.}
% \label{figure:anchor}
% \end{figure}

\subsection{NIRCam Background Level Correction}\label{sec:anchor}

We used the JWST background tool to correct for the differential zero-point variations between exposures. The tool returns background models for a desired wavelength and calendar day. The program was observed in four distinct visits: F300M and F335M were observed on January 9, 11, and 16 of 2023, while F360M was observed on January 10 and 16 of 2023. We found that the average modeled background value at each wavelength is 81.6\ \kjysr\ at 3\um, 73.2\ \kjysr\ at 3.35\um, and 97.6\ \kjysr\ at 3.6\um, and the variation at each wavelength is small but non-negligible over this seven-day time span (at most 1.7\ \kjysr).

Following the method used by the Physics at High Angular resolution in Nearby GalaxieS (PHANGS) collaboration \citep{williams24}, we ensured the pixel values in the NIRCam images were properly zeroed in off-galaxy portions by anchoring to existing wide-area photometry. This was done by matching the background levels of the F300M, F335M, and F360M images to that of the IRAC~3.6\um\ image. Since the field of view is significantly larger for IRAC~3.6\um\ images, the average background level can be measured and removed more reliably. We first convolved the NIRCam images to match the PSF of IRAC~3.6\um\ photometry (FWHM$\sim$1$\farcs$90). Next, we extracted the average surface brightness in concentric annuli of width 2\arcsec\ centered on M101's center from the overlap between the NIRCam and IRAC images, then compared the brightness in each NIRCam filter as a function of IRAC~3.6\um.
%(see Figure~\ref{figure:anchor}). 
We fit a linear relation to the points dimmer than 25~\kjysr\ in both axes and used the intercept to derive an additive offset for the pixel values of each image. After removing background levels using the JWST background tool, the residual offsets subtracted from each image are small: 9.00~\kjysr\ for F300M, 15.82~\kjysr\ for F335M, and $-$5.40~\kjysr\ for F360M.

We validated this anchoring method using the outer edge of our mosaic where there is no discernible galaxy structure. We extracted the brightness in a large aperture (indicated in the bottom panel of Figure~\ref{figure:prettypics}) from the F300M, F335M, F360M, and IRAC~3.6\um\ images. We found that the difference between the NIRCam and IRAC brightnesses in this off-galaxy aperture 
%(shown in the legends of Figure~\ref{figure:anchor})
is equivalent to the intercepts of the fitted linear relations to within 30\% (1.6~\kjysr).

\begin{deluxetable*}{ccccccc}\label{table:onlyone}
\tablecaption{PAH Isolation Parameters for Equation~\ref{eqn:pahisolate}}
\tablehead{\colhead{Parameter} & \colhead{L20} & \colhead{S23} & \colhead{B24$^\star$} & \colhead{Custom B24} & \colhead{T25} & \colhead{This work}}
\startdata
$\alpha$ & 1.00 & 1.68 [1/(1--$\beta$/$B_{\rm PAH}$)] & 1.33 [1/(1--$q\beta$)] & 1.52$\,\pm\,$0.34 & 1.40 [1/(1--$\beta/k$)] & 1.20$\,\pm\,$0.20\\
 \hline
$\beta$ & 0.65 & 0.65 & 0.55 & 0.58$\,\pm\,$0.10 & 0.59 &  0.62$\,\pm\,$0.03\\ 
 \hline
\fpah\ & Spectroscopic & All PAH \threethree\ & All PAH \threethree\ & All PAH \threethree\ & Spectroscopic &  Mean of \\
 & PAH \threethree\ & Correlated & Correlated & Correlated & PAH \threethree\ & L20 \& T25
\enddata
\tablecomments{
$^\star$Derived using F250M in place of F300M \\
(1) $B_{\rm PAH}$ from S23 is the slope of F335M/F300M vs. F360M/F300M for points where F1130W $>$ 10\,\mjysr, $q$ from B24 is defined as F360M = F360M$_{\rm continuum}~+~q~\times~$\fpah, and $k$ from T25 is defined as \fpah\ = $k~\times~$F360M$_{\rm PAH}$.}
\end{deluxetable*}

\subsection{PSF Matching}\label{sec:convolve}

Images from the F300M, F335M, and F360M filters have slightly different PSFs. These PSFs are non-Gaussian and their average width increases for filters at longer wavelengths (FWHM = 0$\farcs$10, 0$\farcs$11, and $0\farcs12$, respectively). To ensure our photometric recovery of the spatial distribution of PAH \threethree\ emission across M101 is not affected by the varying PSF, we convolved the F300M and F335M images to match the PSF of F360M. We used the convolution kernel generation method from the PHANGS collaboration pipeline \citep{phangspipeline} which implements the algorithm of \citet{aniano11}. This method was run with all default parameters, except we chose not to `circularize' the kernels because the NIRCam PSFs are highly azimuthally asymmetric. Instead, we rotated each NIRCam PSF (simulated using the STPSF tool \citep{stpsf}) to match the precise position angle of the telescope at the time of observation. The position angles for the F300M, F335M, and F360M mosaics are approximately equal, differing by at most 0.01~degree. Based on the alignment of point sources in the final convolved images, we find these non-circularized kernels result in a better match between images at different wavelengths.

The same method was also used to produce maps with a resolution of FWHM~=~11$\farcs$2 to facilitate comparisons with the findings of Paper~I where all data were convolved to match PACS~160\um\ imaging. From these images we extracted the regions defined in Paper~I to compare the trend between \pahthree\ and metallicity with the trends seen in the other PAH features.

Finally, we produced a set of all images that are matched to the PSF of MIPS~24\um\ photometry (FWHM~$\sim6\farcs4$). In \S\,\ref{sec:cliff}, these PSF-matched images are used to study radial variations near the apparent `cliff' in the galaxy structure seen at the outermost edge of the NIRCam images in Figure \ref{figure:prettypics}.

\subsection{Isolating the \threethree\ PAH Emission}\label{sec:pahextract}

The F335M filter captures a diversity of emission: the PAH emission feature at \threethree, the aromatic and non-aromatic (aliphatic) features around 3.4\um, the `PAH continuum plateau' \citep{boersma23}, line emission \citep[e.g. Pf$\delta$,][]{lai25}, and the hot dust and starlight continuum emission between $\sim$3.1--3.6\um. We estimate the effect of line contamination in Appendix~\ref{line_contam} and find it is unlikely to significantly alter our results. Reliable comparison with models and prior \spitzer/IRS spectroscopy requires isolating the \threethree\ PAH emission that would have been recovered spectroscopically.  For this we used the co-spatial data in the flanking filters F300M and F360M to estimate the continuum contribution to F335M. However, the short-wavelength end of the F360M filter overlaps with the long-wavelength end of the F335M filter, and as a result captures some of the wing of the \threethree\ PAH feature as well as the aromatic and non-aromatic dust features around 3.4\um\ and the `PAH continuum plateau' that have each been observed to correlate with PAH \threethree\ \citep{lai20, pahfit2025}. To account for the varying contributions in the flanking continuum bands, we investigated a variety of prescriptions that have been calibrated in previous studies.

The first method was presented in \citet[][L20]{lai20} and was calibrated against spectroscopic recovery of the \threethree\ PAH feature from AKARI observations. The method of \citet[][T25]{tarantino25} is based on spectroscopic decomposition of JWST/NIRSpec observations by the PDRs4All collaboration \citep{pahfit2025}. The method presented in \citet[][S23]{sandstrom23} was built on the L20 method by including all emission in F335M that spatially correlates with regions that are also bright in the PAH-dominated F1130W filter. The final method presented in \citet[][B24]{bolatto24} is also tuned to capture all emission that is spatially correlated with dust emission, not exclusively the \threethree\ PAH feature. 
%We do not employ the method presented by \citet{chown25a} because it was calibrated to match observations of the Orion Bar photo-dissociation region environment specifically and our maps cover a much more diverse range of environments.

Based on spectroscopic decomposition of the 3.3--3.5\um\ region of JWST/NIRSPEC observations in M51, the ratio of PAH \threethree\ plus the `PAH continuum' and non-aromatic features at 3.4--3.5\um\ to PAH \threethree\ alone is about 1.7 on average (T. S.-Y. Lai, private communication), similar to the ratio between S23 and L20. This implies that the photometric isolation methods targeting `all dust-correlated emission' (S23 and B24) can be corrected to match the `spectroscopically targeted PAH 3.3\um-only' (L20 and T25). For this reason, we corrected the \fpah\ from the S23 and B24 methods by the ratio between S23 and L20 (1.68) to derive PAH \threethree\ from all four methods. It is important to note that all of these methods are continuum-based and none of them directly control for variations in bright emission lines, which are known to have a significant contribution when PAH \threethree\ is dim \citep{lai25}.

All prescriptions considered in this work can be cast into the following functional form:

\begin{equation}\label{eqn:pahisolate}
    \rm F335M_{PAH} = \alpha~(F335M - [(1-\beta)~F300M + \beta~F360M])
\end{equation}

\noindent each with differing methods for determining the constants $\alpha$ and $\beta$. The constant $\alpha$ quantifies the PAH contribution to the F360M filter and $\beta$ quantifies the slope of the continuum between the three NIRCam filters. As described above, \fpah\ is defined differently in each method, for L20 and T25 it is defined as the PAH \threethree\ emission alone, while the S23 and B24 methods also include in \fpah\ any emission that is dust-correlated. The values used in each prescription are summarized in Table~\ref{table:onlyone}. L20 is the case where the constants of Equation~\ref{eqn:pahisolate} are $\alpha$~=~1 and $\beta$~=~0.65, the latter tuned to best recover the spectroscopically measured \threethree\ power. T25 adopted the constant $\alpha$ = 1.181 and assumed $\beta$ to be the ratio of the pivot wavelength differences of the filters in use ($\frac{3.365-2.996}{3.621-2.996}=0.59$). S23 is the case where $\alpha=1/(1-\beta/B_{\rm PAH})\simeq 1.68$ and $\beta$~=~0.65, giving values which are rescaled from L20 by a fixed factor. B24 adopted parameters that were calibrated to best suit their extraplanar region of interest ($\alpha$ = 1.33 and $\beta$ = 0.55 using F250M instead of F300M), noting that these values are local quantities that vary with PAH properties and continuum color. On this basis, and since we must recalibrate their method to use F300M in place of F250M, we expanded on the method developed by B24, creating a map where the constants $\alpha$ and $\beta$ can vary as a function of radius across M101. However, we find that the derived constants at the center of M101 are statistically equivalent to those at the edge. The values are $\alpha=1.52\,\pm\,0.34$ and $\beta=0.58\,\pm$\,0.10. We use these `custom B24' values for the remainder of this work. We explore the impact of these different photometric recovery methods and detail our adopted prescription in \S\,\ref{sec:pah33_Z}.

Since for this work we want to isolate the \threethree\ PAH feature alone, we adopt the mean of the values of $\alpha$ and $\beta$ used in the spectroscopically targeted methods from L20 and T25, and take half the difference between the two values as an estimate for each parameter's uncertainty, such that $\alpha=1.20\,\pm\,0.20$ and $\beta=0.62\,\pm\,0.03$ (see Table~\ref{table:onlyone}). We propagate the uncertainty between the methods by generating an \fpah\ map from the median and standard deviation of 100 Monte Carlo trials where $\alpha$ and $\beta$ are randomized assuming their uncertainty is Gaussian.

\begin{figure}
\centering
\includegraphics[width=0.9\linewidth, height=0.9\textheight]{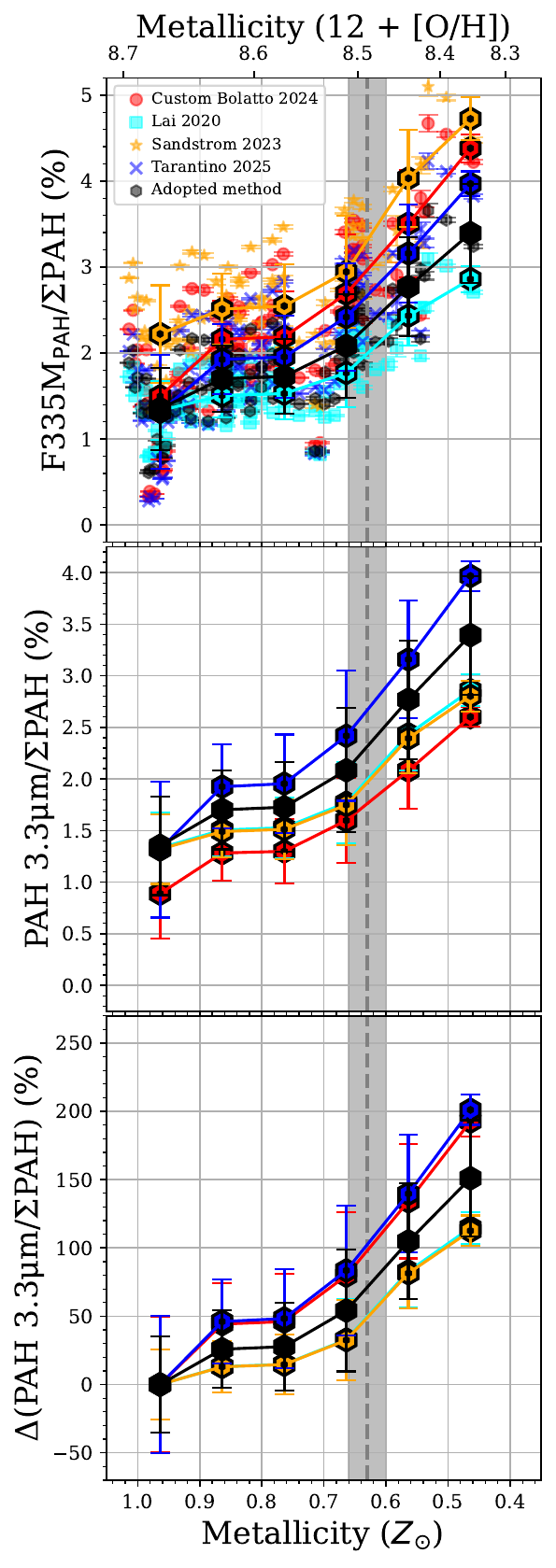}
\caption{\textbf{Top:} \fpah\ as a function of metallicity, as defined in each labeled work. \textbf{Center:} Corrected \pahthree\ as a function of metallicity for the spectroscopic-equivalent \threethree\ PAH feature alone (see \S\,\ref{sec:pah33_Z}). \textbf{Bottom:} fractional change in \pahthree\ with respect to the value at \zsolar.
The dashed gray line indicates the threshold metallicity where \pahtir\ begins to drop. Hexagons represent the median value in six equally-spaced bins from 1.0 to 0.4\,\zsolar\ with errorbars indicating the standard deviation. Note the S23 and L20 trends in the bottom two panels are identical since the methods vary only by a fixed factor.}
\label{figure:pah33_data_pacs160}
\end{figure}

\section{Observational Results}\label{sec:obsresults}

Figure~\ref{figure:prettypics} shows the PAH~\threethree\ map obtained by the continuum subtraction of JWST/NIRCam images adopted in this work. The \threethree\ emission is strongest in the dense regions of the spiral arms, highlighting small \hii\ regions and wispy filaments of ISM between them. The emission has a faint but pervasive background in between the spiral arm structure that is brightest at the center of M101 and decreases with galactocentric radius. At the edge of the image, we note the abrupt `cliff' in emission where the \threethree\ brightness drops significantly. In Appendix~\ref{sec:cliff}, we investigate this cliff by studying radial profiles in emission from bands ranging UV to H\textsc{I} 21 cm. We conclude that the dramatic appearance of this cliff in the NIRCam imaging likely results from the combination of high spatial resolution at the outermost, final spiral arm at this azimuth in M101 and not a specific threshold in metallicity.

In Paper~I, we used the relative shift in power between PAH features longer than 5\um\ to probe changes in the PAH grain population. In this work, we expand on this analysis with the addition of the \threethree\ feature, examining the fractional power relative to total PAH emission as a function of metallicity in Figure~\ref{figure:pah33_data_pacs160}. We derived \fpah\ using the five methods described in \S\,\ref{sec:pahextract}, extracted the apertures defined in Paper~I (green in Figure~\ref{figure:prettypics} bottom), and converted from surface brightness (\mjysr) to integrated intensity (\jdu) for comparison with the \spitzer/IRS results from Paper~I. The points from each method are binned separately, where the connected hexagons shown in Figure~\ref{figure:pah33_data_pacs160} are medians in bins of width 0.1\,\zsolar\ and their uncertainties are the standard deviation in these bins. The JWST/NIRCam footprint covers the full extent of the \spitzer/IRS spectral coverage (out to $\sim0.35$\,\zsolar), however we find the depth is insufficient at large radii to achieve 3$\sigma$ detections of PAH \threethree\ below $\sim0.4$\,\zsolar.

In general, we find the four methods for isolating \fpah\ agree that \fpah/\totpah\ increases with decreasing metallicity, as demonstrated in the top panel of Figure~\ref{figure:pah33_data_pacs160}. \totpah\ is calculated by PAHFIT \citep{pahfit} of the \spitzer/IRS spectroscopy in Paper~I with \fpah\ added. The trend from all methods shows an increase in \fpah/\totpah\ from $1-2$\% at \zsolar\ to $2.5-4.2$\% at 0.45\,\zsolar. The \fpah/\totpah\ ratios from each method vary in overall brightness, with L20 the dimmest and S23 the brightest. The trends from these two methods are offset by a constant factor of 1.68 (as expected since they differ only in the value of $\alpha$ by this factor in Equation~\ref{eqn:pahisolate}). T25 is slightly dimmer than L20 in the center of M101 and brighter at the edge. At a given radius, the T25 image is brighter than the L20 image in bright regions and dimmer than the L20 image in dim regions. This is likely a direct result of the different values of $\beta$ (which quantifies the slope of the continuum between 3 and 3.6\um) used in each. The custom B24 trend falls between the T25 and S23 trends.

\subsection{\pahthree\ vs. Metallicity}\label{sec:pah33_Z}

The top panel of Figure~\ref{figure:pah33_data_pacs160} shows strong agreement in the metallicity trend of \fpah/\totpah\ for maps derived from each method. The main distinction between different versions of these maps is in the absolute intensity of the derived \fpah\ brightness. This distinction can be attributed to differences in assumptions about the shape of the underlying continuum and, most importantly, what emission constitutes \fpah: the L20 and T25 methods are spectroscopically calibrated, tying \fpah\ purely to PAH \threethree\ emission, while S23 and B24 include all emission that is spatially correlated with the PAH emission in F335M. At solar metallicity, the L20, T25, and B24 methods result in \fpah/\totpah\ of about 1--1.5\%, and the S23 method results in \fpah/\totpah\ of about 2\%.

The middle panel of Figure~\ref{figure:pah33_data_pacs160} shows \pahthree\ from each method with the S23 and B24 trends `corrected' by a factor of 1.68 to give spectroscopic equivalent \threethree\ PAH intensity alone. 
%Based on spectroscopic decomposition of the 3.3--3.5\um\ region of JWST/NIRSPEC observations in M51, the ratio of PAH \threethree\ plus the `PAH continuum' and non-aromatic features at 3.4--3.5\um\ to PAH \threethree\ alone is about 1.7 on average (T. S.-Y. Lai, private communication), similar to the ratio between S23 and L20. This indicates that the intensity distinction between photometric isolation methods targeting `all dust-correlated emission' (S23 and B24) and `spectroscopically targeted PAH 3.3\um-only' (L20 and T25) results from different assumptions of what constitutes \fpah. For this reason, we corrected S23 and B24 by the ratio between S23 and L20 (1.68). 
After dividing this factor from the S23 and B24 maps to derive the \threethree\ PAH intensity exclusively, we found the S23 map is equal to the L20 map, and B24 is the dimmest map with \pahthree~$\sim$~0.75\% at \zsolar. This correction reveals that the spatial variations in the recovered PAH \threethree\ intensity are all equivalent between the methods to within the 1$\sigma$ scatter around the median, as shown in the middle panel of Figure~\ref{figure:pah33_data_pacs160}.

In the bottom panel of Figure~\ref{figure:pah33_data_pacs160} we normalized each trend to its value at solar metallicity. Throughout the range of our NIRCam coverage (from \zsolar\ to $\sim$0.4\,\zsolar) the L20 and S23 maps show a fractional increase in \pahthree\ of about 120\%, while the B24 and T25 maps show an increase of about 200\%. This distinction is a result of differences in the value of $\beta$ used in each method. The \pahthree\ trend with metallicity from the isolation method adopted in this work shows a fractional increase of about 150~$\pm$~50\% (i.e.,~a~factor~of~$\sim2.5$).

%Since for this work we want to isolate the \threethree\ PAH feature alone, we adopt the mean of the values of $\alpha$ and $\beta$ used in the spectroscopically targeted methods from L20 and T25, and take half the difference between the two values as an estimate for each parameter's uncertainty, such that $\alpha=1.20\,\pm\,0.20$ and $\beta=0.62\,\pm\,0.03$ (see Table~\ref{table:onlyone}). We propagate the uncertainty between the methods by generating an \fpah\ map from the median and standard deviation of 50 Monte Carlo trials where $\alpha$ and $\beta$ are randomized assuming their uncertainty is Gaussian.

%As metallicity drops from solar to the threshold value \zthresh~$\sim~$0.63\,\zsolar, \pahthree\ shows a increase of about 30\% in maps made with the L20 and S23 methods. The trend from the B24 and T25 maps shows an increase in \pahthree\ of about 80\% between \zsolar\ and \zthresh. 
%Relative to the value at \zthresh, at the lowest metallicity where we have significant PAH detections ($\sim$0.45\,\zsolar) \pahthree\ shows a dramatic increase of approximately 100\% from the L20 and S23 maps, and approximately 120\% from the B24 and T25 maps. By design, the trend considered in this work falls between that of all four methods, rising 55\% from \zsolar\ to \zthresh\ and another 110\% from \zthresh\ to 0.45\,\zsolar.

\section{Modeled PAH Analysis}\label{sec:dvresults}

\subsection{\citet{whitcomb24} PAH Models}

In Paper~I, we used the D21 framework to explore three physically motivated models that could potentially reproduce the observed PZR trends. The D21 models allow us to generate theoretical infrared spectra for an input grain size distribution, grain ionization distribution, and illuminating spectral energy distribution (SED) shape and intensity. The first model we tested was a radiation-hardness-only model in which the PAH population size, charge, and abundance are left fixed and the hardness of the illuminating radiation field increases toward lower metallicity. This model resulted in the correct direction of the shift in PAH power to shorter wavelength features. However, we found that the magnitude of this shift was grossly insufficient compared to the observations even in the most extreme possible radiation case. Next, we used a model where the decrease in overall PAH abundance (\qpah) was a result of photo-destruction of PAH grains from the smallest up. We found that to reproduce the steep decrease in \qpah, it was necessary to destroy all PAHs with fewer than $\sim400$ carbon atoms. In doing so, the resulting band ratios exhibited a trend opposite to that found from the observations: the power shifted away from short-wavelength features and toward long-wavelength features.

Finally, the best-fitting model invoked a form of inhibited grain growth, in which the average PAH grain size decreases with metallicity and the overall mass of the PAH grains declines below \zthresh. This model successfully reproduces both the observed band ratios and the behavior of \qpah\ below \zthresh. The model was calibrated to match the trends seen in all major PAH features longer than 5\um, but also predicted the behavior of \pahthree\ with decreasing metallicity. Since the relative number of the smallest PAH grains increases significantly in the inhibited growth model, we predicted \pahthree\ would more than triple from \zsolar\ to $\sim$0.4\,\zsolar.

%Here, the observations show a significant rise in \pahthree\ at low metallicity, but the trend adopted in this work does not show a rise as steep as the model in Paper~I predicted.

\begin{figure*}
\centering
\includegraphics[width=\linewidth]{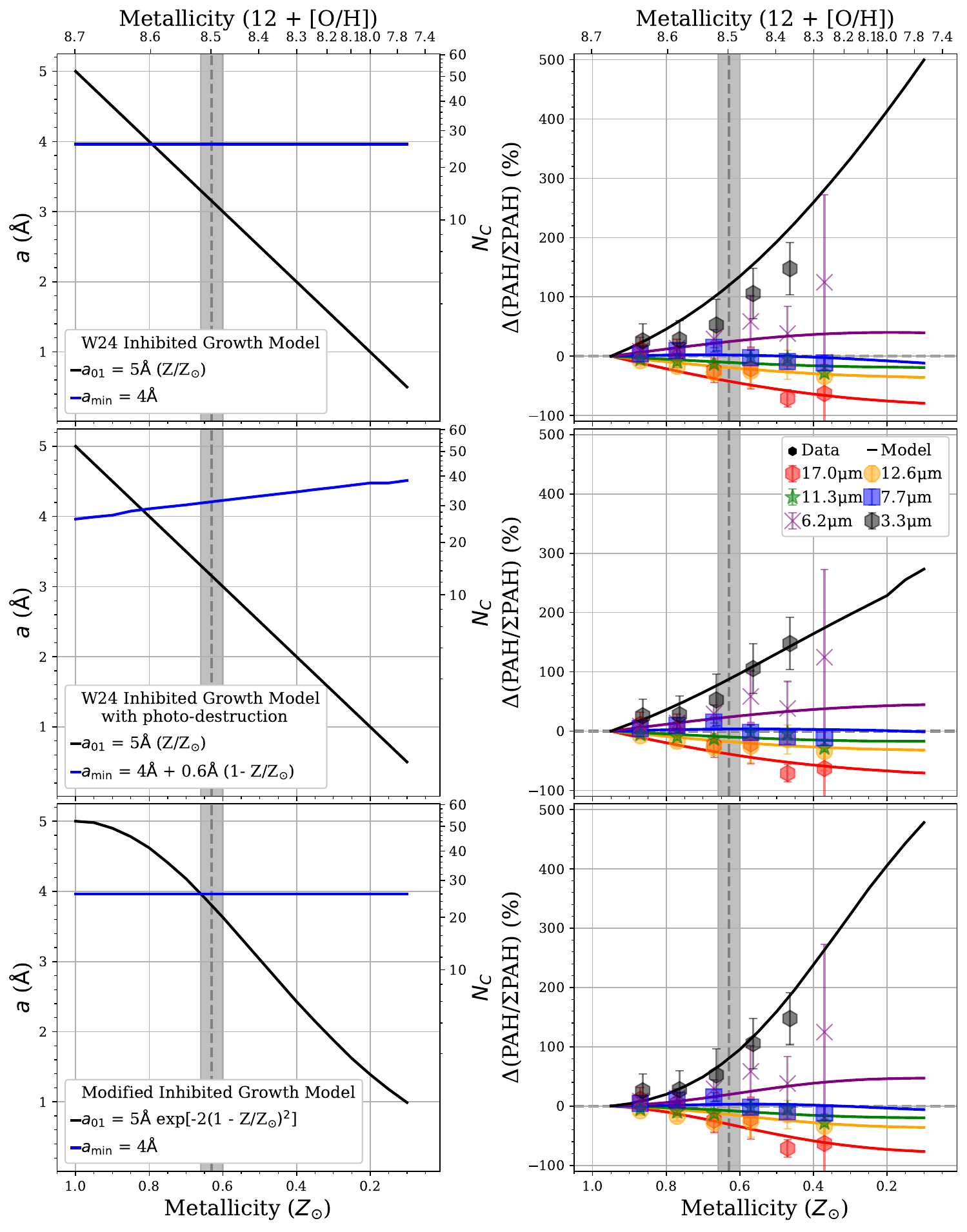}
\caption{(Left) modeled metallicity-evolution of the characteristic grain size \aone\ (black) and the minimum grain size \amin\ (blue) for each model (see \S\,\ref{sec:dvresults}). (Right) fractional change in all major PAH features with predictions from various inhibited growth models indicated by solid lines: linear inhibited growth from Paper~I with fixed standard \amin\ = 4$\rm \AA$ (\carbons\ = 26) \textbf{(top)}, linear inhibited growth with \amin\ increasing linearly up to 4.6$\rm \AA$ (\carbons\ = 40) \textbf{(center)}, and a modified inhibited growth model with fixed standard \amin\ \textbf{(bottom)}. The dashed gray line indicates the threshold metallicity from Paper~I ($\sim$0.63\,\zsolar).}
\label{figure:shredder}
\end{figure*}

The top panel of Figure~\ref{figure:shredder} shows the \pahthree\ values from the bottom panel of Figure~\ref{figure:pah33_data_pacs160}, plotted with the other fractional PAH changes and W24 model predictions from Paper~I (where \aone($Z$)~=~$5\rm \AA$~($Z$/\zsolar)). The agreement between the model prediction and the data is quite good, although the model slightly overestimates the change in \pahthree\ by $\sim20-70\%$ at 0.45\,\zsolar.

Emission from the smallest PAH grains is uniquely sensitive to even modest amounts of photo-destruction. While the results confirm that enhanced destruction cannot explain the broad trends in the PZR, we modify the W24 inhibited growth model to explore how much photo-destruction can be accommodated at the smallest end of the grain size distribution.  In addition, we consider an alternative functional form for the variation of the characteristic PAH grain size \aone($Z$) which is sub-linear at metallicities higher than \zthresh\ and super-linear at lower metallicities. All models use the modified solar neighborhood interstellar radiation field SED \citep{mmp83, D21} with intensity $U=1$ and the standard ionization distribution as a function of grain size.

\subsection{Inhibited Growth with Limited Photo-destruction}\label{sec:IG1}

As shown in Paper~I, photo-destruction of PAH grains from the smallest up cannot be invoked to explain the observed decrease in the mass fraction of PAH-emitting grains (\qpah). However, without observations of the \threethree\ feature, we were unable to rule out a modest amount of photo-destruction affecting only the very smallest PAHs.

To explore the amount of photo-destruction that could be consistent with observations, we expanded on the inhibited growth model from Paper~I (where the characteristic grain size \aone\ decreases linearly with metallicity), adding a modest amount of small grain destruction to compare to the observed shift in \pahthree.

We implemented photo-destruction of small grains in our inhibited growth model by increasing the minimum grain size \amin\ with a simple linear relation:

\begin{equation}\label{eqn:amin}
    a_{\rm min} = a_{\rm min,0} + \Delta a \left(1-\frac{Z}{~Z_\odot}\right)
\end{equation}

\noindent where $a_{\rm min, 0}$ is the initial minimum size used by D21 ($4\,\rm\AA$ or \carbons~=~26) and $\Delta a$ is an adjustable parameter. All grains below \amin\ were removed from the distribution before the resulting spectrum was calculated. We found that with the functional form assumed in Equation~\ref{eqn:amin} and the W24 linear inhibited growth model, $\Delta a$ can be any value between 0.2$\,\rm\AA$ -- 1.1$\,\rm\AA$ and remain consistent with the \pahthree\ observations to within 1$\sigma$. This range corresponds to a maximum \amin\ of 4.2$\,\rm\AA-5.1\,\rm\AA$ (\carbons\:$=\:30-55$), which is consistent with previous theoretical works that found that photo-destruction by FUV might be able to remove grains with up to \carbons\:$\sim40-50$ (i.e. $a=4.6\,\rm\AA-4.9\,\rm\AA$), depending on the environment \citep{guha89, allain96}. The best agreement with the \pahthree\ observations is when $\Delta a$ = 0.6$~\pm~0.2\,\rm\AA$.

The result of this model is shown in the center panel of Figure~\ref{figure:shredder}. Compared to the model from Paper~I shown in the top panel, the model with limited photo-destruction better matches the observed trend in \pahthree\ as a function of decreasing metallicity without affecting the trends in the other longer-wavelength PAH features.

\subsection{Inhibited Growth: Modified Functional Form}\label{sec:IG2}

Finally, we explored a possible alternative functional form for the inhibited growth model that could better match the observed \pahthree\ variations. In Paper~I, we assumed a linear functional form for \aone, which characterizes the location of the small-radius peak in the grain size distribution, such that \aone($Z$)~=~5$\rm \AA$~($Z$/\zsolar). The linear \aone($Z$) model mildly overestimates measurements at high metallicities and underestimates at low metallicities for many features.  Given that the grain size distribution modeled in D21 is log-normal in $a$, it may be more natural to assume that $\ln$(\aone) is the parameter that should vary with metallicity. This change also affects all of the observed fractional PAH feature variations with metallicity.

We assumed a simple functional form such that:

\begin{equation}
    \mathrm{ln}\left(\frac{a_{01}}{a_{01,0}}\right) = -b \left(1- \frac{Z}{~Z_\odot}\right)^c \quad \mathrm{for}~Z<Z_{\odot}
\end{equation}

\noindent where \aone\ characterizes the location of the small-radius peak in the D21 size distribution of carbonaceous grains, $a_{01,0}$ is the initial value of \aone\ at $Z=Z_\odot$, and $b$ and $c$ are constants that result in the closest agreement with the observations. As in the linear inhibited growth model, we adopt the standard \amin\ = $4\,\rm\AA$ and an initial value of \aone\ equal to the largest value of \aone\ considered in D21 ($a_{01,0}=5$\AA). With insufficient data at metallicities $>$\zsolar, we assume \aone\ remains constant at $a_{01,0}$ for super solar metallicities. The best agreement with the observations is achieved with constants $b=2$ and $c=2$, so the final functional form of $\aone(Z)$ is a semi-Gaussian with mean \zsolar\ for $Z<$\zsolar.

% The resulting form of the grain size distribution as a function of metallicity is then:

% \begin{equation}
%     \left(\frac{dn_{\rm PAH}}{~da}\right) = \frac{n_{\rm H}}{a} \left(B_1~\rm exp\left[\frac{(Z/Z_\odot)^2~-~\rm ln(a/5\rm \AA)}{~2\sigma^2} \right] \right)\quad \mathrm{for}~Z<Z_{\odot}
% \end{equation}

The PZR trend produced by this model is shown in the bottom right panel of Figure~\ref{figure:shredder}.
The change in the functional form of the inhibited growth model results in excellent agreement with the observed trend in \pahthree\ without the need to invoke photo-destruction of small grains. The agreement is also improved for all other fractional PAH feature trends at metallicities greater than the threshold value \zthresh\,$\,\approx\,0.63$\,\zsolar. Below \zthresh, this modified inhibited growth model with semi-Gaussian functional form is approximately equivalent to the original linear model for all fractional PAH feature trends longer than 5\um. The two models become distinct for the \pahthree\ trend, where the agreement is significantly better for the modified model compared to the original model. 

%This agreement could be improved or worsened depending on the choice of continuum subtraction method since our adopted method represents an average of each.

\begin{figure*}
\includegraphics[width=0.99\linewidth, height=0.75\textheight]{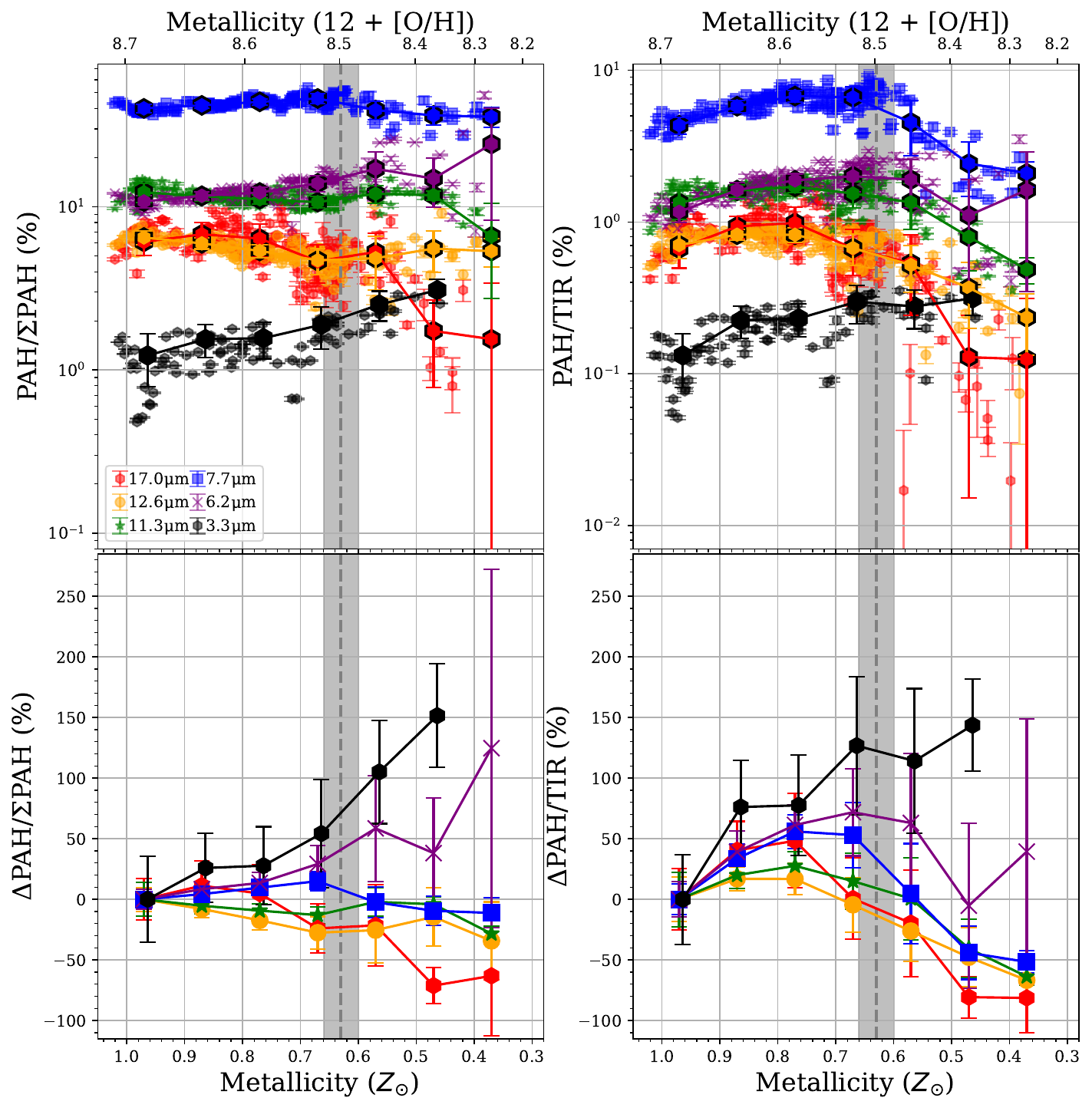}
\caption{\textbf{Top left:} PAH to Total PAH ratios as a function of metallicity, \textbf{top right:} PAH to TIR ratios as a function of metallicity, \textbf{bottom left:} fractional change in PAH to Total PAH ratios relative to the value at \zsolar, \textbf{bottom right:} fractional change in PAH to TIR ratios relative to the value at \zsolar.
The dashed gray line indicates the threshold metallicity where \pahtir\ begins to drop. Hexagons represent the median value in bins of width 0.1\,\zsolar\ with errorbars indicating the standard deviation. Note the trend for all PAH features except \threethree\ is taken from Paper~I and these include data from NGC~628 and NGC~2403 in addition to M101.}
\label{figure:pahtir_data_pacs160}
\end{figure*}

\section{Discussion}\label{sec:disc}

The results on the observed PAH population from this work combined with those from Paper~I provide insights on the equilibrium between the competing effects of grain growth through accretion and coagulation, and grain destruction through shattering and photo-processing. The observations indicate that an inversion of typical dust processing occurs for the PAH population. Processing is typically assumed to be stronger for the smallest of the large grains \citep{madden06, gordon08, micel10a, micel10b}, but our results suggest, surprisingly, that the smallest of the small grains are found in \emph{greater} abundance relative to their larger cousins as metallicity decreases. The agreement between our models and the observations suggests that the processes that determine the PAH population from the bottom up, i.e. the accretion of carbon atoms and coagulation of PAH precursors, are being inhibited as metallicity decreases.

Figure~\ref{figure:pahtir_data_pacs160} summarizes our observations of the PAH-metallicity trends in all major PAH features in the nearby spiral galaxies M101, NGC~628, and NGC~2403 (with \threethree\ data only from M101). The left panels show how power shifts among the PAH features away from longer wavelengths toward shorter wavelengths. Of particular note, we found that below 0.5\,\zsolar\ the fractional contribution of the \threethree\ feature to the total PAH power surpasses that of the 17\um\ feature. If the trend continues, we would predict that the fractional contribution of the \threethree\ feature would begin to surpass that of the 12.6\um\ feature as well at metallicities below $\sim0.35$\,\zsolar.
The ratio of the strength of PAH bands longer than 12\um\ to the 3.3\um\ band place strong constraints on this class of models.  Future work exploring such band ratios as a function of metallicity over the key metallicity range 0.2–0.4~\zsolar\ will be very informative.
This highlights the pivotal role that short-wavelength PAH features can play in studies of low-metallicity galaxies and/or high-redshift galaxies where metallicities may naturally be lower on average.

\subsection{Interpretation of the Inhibited Growth Model}

The inhibited growth model from Paper~I results in an excellent first-order approximation of the observed fractional PAH band strengths as a function of metallicity. The model predicted a steep increase in the \pahthree\ ratio as metallicity decreases below \zsolar, and indeed this is what JWST observations show. The magnitude of the increase is steeper in the model than in the observations, but only by slightly more than the 1$\sigma$ uncertainty. Either of the two modifications to this model presented in \S\,\ref{sec:IG1} (limited photo-destruction) and \S\,\ref{sec:IG2} (semi-Gaussian functional form) can be invoked to explain this discrepancy. The two modified models only become significantly distinguishable at metallicities below $\sim$20\%, so observations of galaxies with this metallicity and lower may allow them to be differentiated. Future studies that explore the photon effects on PAHs in environments with higher radiation field intensities can further constrain the hybrid and pure inhibited growth models presented in this work.

If PAH (re-)formation occurs predominantly from coagulation of PAH precursors and/or accretion of carbon atoms --- i.e., from the bottom up \citep{wenzel25, lee25, 30dor, zhang25} --- inhibited growth of larger PAH grains at low metallicities is a reasonable explanation for the observed PZR trends. In this case, the mechanism(s) responsible for the growth of small grains into larger grains must be inhibited as metallicity declines. In general, at lower metallicity there is less dust to attenuate photons in the ambient radiation field, causing the radiation field to be harder and more intense than in high-metallicity environments. If individual carbon atoms or other PAH precursors are accreted and grow on the surfaces of larger grains, their reduced availability and longer accretion timescales at lower metallicity could explain the overall shift toward smaller grain sizes, although the increased intensity and hardness of radiation could also play a role in inhibiting growth. The result of inhibited growth would be a decrease in the amount of PAHs of all sizes, but preferentially affecting larger grains. This interpretation is consistent with previous theoretical works on the metallicity dependence of grain growth \citep{zhukovska08, zhukovska13}

As discussed in Paper I, any physical mechanism that results in smaller overall PAH sizes could produce variations in the PAH spectrum similar to those observed in this study. For example, selective processing of the largest PAH grains as metallicity decreases would enhance the relative abundance of small grains and result in a similar shift in emitted power toward shorter wavelength PAH features. Photo-processing is likely to be size dependent, with the greatest impact on the smallest grains \citep{hunt10, D21}. However, it remains possible that there are additional mechanisms that preferentially shield very small PAHs from photo-destruction, such as recurrent fluorescence \citep{wittlai20, lai23}.  When combined with broader grain processing effects, these protection mechanisms could result in an effective net decrease in the average PAH size. Future modeling will be required to determine the significance of these effects.

\begin{figure}
\includegraphics[width=0.99\linewidth]{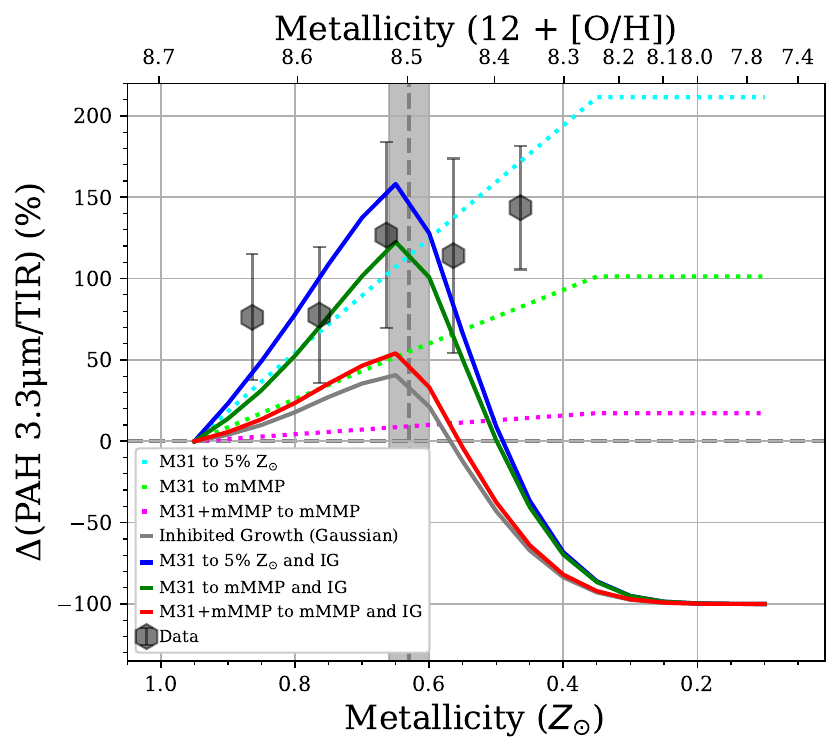}
\caption{Fractional change in the ratio of PAH \threethree\ to TIR relative to the value at \zsolar\ with D21 model lines overplotted. Dotted lines indicate models where only the incident radiation field changes with metallicity, and solid lines indicate these models combined with the inhibited growth model from \S\,\ref{sec:IG2} shown in solid gray.
The dashed gray line indicates the threshold metallicity where \pahtir\ begins to drop.}
\label{figure:threetir}
\end{figure}

\subsection{Mass Fraction of Ultra-Small Grains Rises}\label{section:threetir}

The right panels of Figure~\ref{figure:pahtir_data_pacs160} show that \threethree\ exhibits a surprising trend not seen in any other PAH feature. The top panel makes clear that all other major PAH features begin to decrease in strength relative to total dust luminosity (TIR) below the threshold metallicity. Unusually, although the uncertainties are large at these lower metallicities, the fractional power of \threethree\ relative to all dust emission appears to continue rising towards lower metallicity.  Using TIR as an approximate proxy for the total dust mass and PAH \threethree\ as a proxy for the mass of ultra-small carbonaceous grains with fewer than $\sim$100 carbon atoms, this may imply that the mass fraction of ultra-small grains relative to all dust actually \emph{increases} as metallicity declines.

An additional consideration is that ultra-small grains are far more sensitive to changes in the hardness of the illuminating radiation field than the larger grains which contribute the bulk of the dust luminosity.  
Using the D21 models, we find in the most extreme case where the radiation field SED changes from that of the M31 bulge to a 5\%\,\zsolar\ 10 Myr old stellar population results in a predicted increase in PAH \threethree/TIR of about 210\%, as shown in the upper dashed curve in Figure~\ref{figure:threetir}.  A more realistic model of how the radiation field varies across M101 approximates the radiation field at the center as a 30\%:70\% mix of the M31 SED and the mMMP SED (based on our analysis in Section\,4.1.1 of Paper~I) and the radiation field in the outskirts as a pure mMMP SED, with a linear interpolation between. In this case, we find PAH \threethree/TIR increases only $\sim$17\% from this change in radiation hardness alone. For an intermediate model that ranges from the M31 bulge SED at the center to the mMMP SED at the edge, PAH \threethree/TIR increases by about 100\%.

However, each of these models assumes that the PAH grain size distribution remains fixed, which is at odds with the rapid drop in total PAH power. The grain size evolution of the best-fitting inhibited growth model from \S\,\ref{sec:IG2} predicts PAH \threethree/TIR peaking at \zthresh, then falling at lower metallicity as the total PAH mass rapidly declines. Combining the inhibited growth model with each of the three varying radiation field models improves the agreement with measurements, but none can reproduce the excess PAH \threethree/TIR ratio at the lowest metallicity bin at 0.45\,\zsolar. Since the PAHs that dominate emission at \threethree\ represent a very small fraction of the total PAH mass, this discrepancy could be a result of the very smallest grains deviating from the overall size evolution process. Any mechanism that enhances the survivability of ultra-small PAHs, e.g. recurrent fluorescence \citep{wittlai20, lai23}, could protect these grains from the broader decline in PAH abundance, bringing the model into closer agreement with the observations.  Alternatively, the ionization distribution may be changing with metallicity. If grains are becoming more neutral on average as metallicity decreases, this would increase the brightness of \threethree\ relative to the other PAH features that originate more preferentially from ionized PAHs. As these interpretations hinge on a single point at $\sim$40\% \zsolar, further observations in the critical metallicity regime 20~--~50\%\,\zsolar\ may shed more light on this phenomenon.

\subsection{Constraining Small Grain photo-destruction}

In Paper~I we found that photo-destruction of PAH grains cannot be a primary driver of the observed PZR trends.  However, this effect may still influence the metallicity trend of PAH \threethree\ alone, as this feature is emitted primarily by the smallest PAH grains that are expected to be the most sensitive to small changes in the minimum surviving grain size. In \S\,\ref{sec:dvresults}, we found the inhibited growth model from Paper~I predicts slightly too steep of an increase in \pahthree\ as metallicity decreases. We then showed in \S\,\ref{sec:IG1} that including a limited amount of small grain photo-destruction can bring the model to match the observations more closely without affecting the trends of PAH features longer than 5\um. 

However, invoking photo-destruction is a double-edged sword. If we assert that small grains are being destroyed or de-aromatized as metallicity decreases because the hardness of the radiation field is increasing, this same increase in hardness would cause the surviving PAHs of all sizes to reach higher stochastic temperatures and therefore emit more strongly at shorter wavelengths, counteracting some of the decrease in \pahthree\ \citep{baron24, baron25}. As shown in Paper~I and D21, if the grain size distribution is left unchanged and only the hardness of the illuminating radiation field is increased, the result is a small shift in PAH power from long-wavelength features to short-wavelength features. This is consistent with the findings of \citet{dale25} where PAH feature ratios observed in star clusters show little variation as a function of cluster age, and with \citet{baron24} where, at high metallicities, most of the change in the ratio of PAH 11.3\um\ to PAH 7.7\um\ could be accounted for by changes in the hardness of the illuminating radiation field. In the case where the radiation field changes from that of M31's bulge at \zsolar\ to a 5\%\,\zsolar\ low-metallicity starburst SED, \pahthree\ is expected to increase by at most $\sim$40\%. We found that even this extreme hardness effect was entirely insufficient to explain the observed band ratios. This remains true for \pahthree. But increased photon hardness could counteract a portion of any effect of photo-destruction on the \pahthree\ trend.

In \S\,\ref{sec:IG1} and Figure~\ref{figure:shredder}, we show that the maximum change in \amin\ that is consistent with the observations decreases \pahthree\ by about 100\% at 0.45\,\zsolar, while the most extreme case of radiation hardness variation only increases \pahthree\ by about 40\% at 0.45\,\zsolar. The radiation hardness model is based on more extreme assumptions than are realistic for the interstellar conditions found in M101, so future models incorporating realistic photo-destruction vs.\ radiative heating scenarios could further explore this tension. Regardless, even if we assume the maximum effect of radiation hardness, this would only increase our estimate of the minimum surviving grain size \amin\ to 5.4\,\AA (\carbons\,$\sim65$). Therefore, we conclude that photo-destruction is unlikely to have a significant bulk effect on the PAH population in the ISM of spiral galaxies like M101 on the physical scales considered in this work.

Since the carriers of PAH \threethree\ are thought to be sensitive to photo-destruction, we can use the rise in \pahthree\ to place an upper limit on the allowable amount of photo-destruction. We found that with the linear inhibited model from Paper~I, the lower size cutoff \amin\ can rise to at most 5.1\,\AA (\carbons$\:\sim55$) and still match the observations within 1$\sigma$.  In the context of the modified inhibited growth model from \S\,\ref{sec:IG2}, \amin\ can rise to at most 4.57$\rm\AA$ (\carbons$\:\sim40$).  Together these tightly constrain any role for direct photo-destruction of the smallest PAH grains to a very limited range, assuming the absorption cross sections, ionization distribution, and size distribution used in the D21 theoretical framework.

\subsection{Projections for Ultra-Low Metallicity Regime}

Recent observations of low-metallicity and dwarf galaxies exhibit a deficit of PAHs similar to that seen in the nearby spiral galaxies studied in our sample \citep{chown25b, 30dor}. The strong agreement between the observations and our inhibited growth models allows us to forecast future observations of PAH emission in yet lower metallicity environments. \citet{lai25} found the average between upper and lower limits on \pahthree\ ranges from 5--10\% using JWST/NIRSpec and MIRI spectroscopy of the compact dwarf starburst galaxy II~Zw~40 at $\sim$25\%\,\zsolar. At that abundance, the photo-destruction modified model predicts PAH \threethree\ will constitute $\sim$4\% of all PAH emission and the semi-Gaussian modified model predicts it will constitute $\sim$5.5\%. However, the radiation environment is much more extreme in II~Zw~40 than in the outskirts of M101, with a significantly harder starlight SED at much higher intensity. If photo-destruction is modest, each of these aspects will act to increase the observed PAH \threethree\ luminosity. Based on the results of our radiation hardness model, we would expect these effects to increase the observed PAH \threethree\ luminosity above what the model would predict from metallicity alone to $\sim$5.6\% of all PAH luminosity with the photo-destruction modified model and $\sim$7.7\% with the semi-Gaussian modified model. This estimate does not include the effects of the increased radiation intensity, but it assumes the most extreme possible disparity between the hardness of the radiation in the two galaxies. Therefore at 25\%\,\zsolar, both modified inhibited growth models are consistent with the observations from \citet{lai25}.

At 10\%\,\zsolar\ the two modified inhibited grain growth models will be more easily distinguishable. In regions where the radiation environment is similar to the mMMP SED at an intensity of $U=1$, the photo-destruction modified model predicts PAH \threethree\ will constitute $\sim$4.5\% of all remaining PAH emission and the semi-Gaussian modified model predicts it will constitute $\sim$6.9\%. If we assume a more intense and harder radiation field (similar to that of II~Zw 40) at this metallicity, predicted PAH \threethree\ power rises to $\sim$6.3\% or $\sim$9.7\% of all PAH emission for the respective modified models. This range is in agreement with the findings of \citet{tarantino25}, who derive an upper limit on \pahthree\ of 7.7\% of the total PAH power in the dwarf galaxy Sextans A, where the metallicity is $\sim$7\%\,\zsolar. We do note that, based on extrapolation of our observed PZR trends, at this metallicity the total PAH emission is projected to represent well less than 1\% of the total infrared luminosity, so care must be taken to accurately recover individual PAH feature power.

\subsubsection{High-redshift Applications}

Recent observations of galaxies at $z\sim1-1.5$ where metallicities are lower at a given stellar mass have hinted at broad PAH deficits, potentially similar to those observed in the local Universe \citep{shivaei24, lyu25}.  The properties and emergence timeline of dust at earlier epochs are not well known. Due to the sensitivity of PAH grains to the metal content and heating conditions within galaxies, and the high luminosity of PAH emission, they offer strong potential for studying early dust in detail.  Simulations show that a future background-limited far-infrared observatory with spectroscopic capabilities \citep[e.g. PRIMA,][ see whitepaper therein by Donnelly et al.]{glenn25, primaGO1} could offer an effective way to recover dust conditions and metal content in the early Universe out to $z\sim6$ or beyond \citep[see also][]{burgarella25,yoon25}.

\section{Conclusions}\label{sec:concl}

We explored the behavior of the smallest PAH grains with metallicity in a radial strip map of the \threethree\ PAH feature extending across the nearby spiral galaxy M101, which has a steep and well-characterized metallicity gradient. We found the fractional contribution of \threethree\ to the total PAH luminosity increases significantly, rising from $\sim$1\% at \zsolar\ to $\sim$3\% at 0.4\,\zsolar. In conjunction with the findings of Paper~I covering all the PAH features longer than 5\um, we found that the changes in PAH \threethree\ emission are consistent with slightly updated versions of prior models of inhibited grain growth, and that these models agree with recent observations at even lower metallicity. The main findings of this work are as follows.

\begin{itemize}

    \item The spatial distribution of PAH \threethree\ emission at $\sim3.9$\,pc resolution exhibits a sharp cliff at the outer edge of M101 (see Figures~\ref{figure:prettypics} and \ref{figure:radprofs}). Since stellar and dust continuum tracers are agnostic of the cliff, we hypothesize that its appearance in the PAH \threethree\ map is simply a confluence of decreased PAH abundance and low ISM density corresponding to the outermost spiral arm at this azimuth in M101, rather than a special threshold in metallicity.

    \item The fractional power \pahthree\ increases significantly as metallicity drops from \zsolar\ to 0.4\,\zsolar, as predicted by the inhibited grain growth model of \citet{whitcomb24}. The modeled increase is steeper than the observations show, but only slightly above the 1$\sigma$ systematic and propagated uncertainties (see Figure~\ref{figure:shredder}).

    \item The inhibited growth model can be refined to better match the observed trend in \pahthree\ by introducing a limited amount of photo-destruction which removes grains with fewer than 40 carbon atoms at the lowest metallicities (see \S\,\ref{sec:IG1}). These models become inconsistent  with the observed trend if the minimum grain size cutoff exceeds $\sim$55 carbon atoms, sharply limiting the role of photo-destruction in the PAH--metallicity relation.

    \item An alternate inhibited growth model which changes the underlying functional form of grain size evolution with metallicity from linear to semi-Gaussian provides a slightly better match to the observed trends for all PAH features, without invoking a distinct physical mechanism (see \S\,\ref{sec:IG2}).
    
    \item The ratio of PAH \threethree\ to total infrared luminosity actually \emph{increases} as metallicity drops, potentially indicating that the mass fraction of ultra-small grains relative to the full dust grain population rises at low metallicity in a way that deviates from the overall PAH nanoparticle deficit (see \S\,\ref{section:threetir}).  Enhanced heating by harder radiation at low metallicity may also play a role.

    \item Extrapolating our two inhibited grain growth models to lower metallicities, we predict that PAH \threethree\ will constitute $\sim$$4-5.5\%$ of \totpah\ at 25\%\,\zsolar\ and $\sim$$4.5-6.9\%$ at 10\%\,\zsolar. Below $\sim$20\%\,\zsolar, these two models are readily distinguishable, so future observations in this range will help further constrain the physical processes impacting the smallest PAH grains.
    
\end{itemize}

\begin{acknowledgements}

This work is based in part on observations made with the NASA/ESA/CSA James Webb Space Telescope. The JWST data were obtained from the Mikulski Archive for Space Telescopes at the Space Telescope Science Institute, which is operated by the Association of Universities for Research in Astronomy, Inc., under NASA contract NAS 5-03127 for JWST. The specific observations analyzed can be accessed via \dataset[doi: 10.17909/6f77-dn28]{https://doi.org/10.17909/6f77-dn28}. These observations are associated with program \#2452. JDTS and CMW acknowledge NASA/STScI support through contract JWST-GO-02452.001-A. This research was carried out in part at the Jet Propulsion Laboratory, California Institute of Technology, under a contract with the National Aeronautics and Space Administration (80NM0018D0004). This research has made use of NASA's Astrophysics Data System. This work has made use of data from the European Space Agency mission Gaia (\url{https://www.cosmos.esa.int/gaia}), processed by the Gaia Data Processing and Analysis Consortium (DPAC, \url{https://www.cosmos.esa.int/web/gaia/dpac/consortium}). Funding for the DPAC has been provided by national institutions, in particular the institutions participating in the Gaia Multilateral Agreement. This research has made use of the NASA/IPAC Infrared Science Archive, which is funded by NASA and operated by the California Institute of Technology.

\facility{GALEX, Herschel, IRSA, JWST, Spitzer, VLA}

\end{acknowledgements}

\begin{appendix}

\section{The effect of line contamination on photometric recovery of PAH \threethree}\label{line_contam}

We used the following method to quantify the effect of line contamination on our calculated PAH 3.3\um\ band luminosity. We used H$\alpha$ measurements from \citet{fira} in three 10\arcsec\ radius apertures at the center and edge of M101 which are covered by our NIRCam footprint. To estimate the Pf$\delta$ emission in these apertures, we assumed case B applies, such that the ratio of H$\alpha$/Pf$\delta\sim1200$ \citep{storey95}. The apertures are centered on bright \hii\ regions, and so provide an idea of an upper limit of Pf$\delta$ contamination since the regions of M101 probed in this work by the apertures shown in Figure~\ref{figure:prettypics} are of primarily diffuse ISM with only a few small \hii\ regions scattered throughout. We extracted the Lamarche et al. apertures from our adopted PAH 3.3\um\ map, and we find that the estimated ratio of Pf$\delta$ to the photometrically determined PAH 3.3\um\ intensity is $\sim$0.1\% for the two fainter central regions and at most 1.5\% in the bright \hii\ region NGC5455 that falls in our NIRCam strip, but outside the Spitzer/IRS coverage  (i.e. not included in the PZR analysis).

As shown in Figure~7 of \citet{lai25} the line contamination in the flanking continuum bands F300M and F360M is likely a larger source of error than direct Pf$\delta$ contamination in F335M itself. At 20\,pc scales on the bright \hii\ region studied by \citet{lai25}, the line strengths in F300M and F360M act to artificially increase the estimated continuum contribution to F335M, resulting in an underestimate of the true PAH 3.3\um\ integrated intensity by a factor of $\sim$2–4. In M101, we are probing $\sim350\,$pc scales at an azimuth of the galaxy specifically chosen to have as few \hii\ regions as possible, so we do not expect to underestimate PAH 3.3\um\ by such a significant factor as found in \citet{lai25}. Since we expect the effect of line contamination to increase radially (and with decreasing metallicity) in M101, the net result would be a slightly faster rise in Figure~\ref{figure:pah33_data_pacs160} of the fractional PAH 3.3\um\ power as a function of metallicity. Furthermore, any effects of line contamination would likely modestly increase the true trend in the PAH 3.3\um\ feature strength with metallicity and would not significantly alter our interpretations of those trends.

\begin{figure*}
\centering
\includegraphics[width=0.99\linewidth]{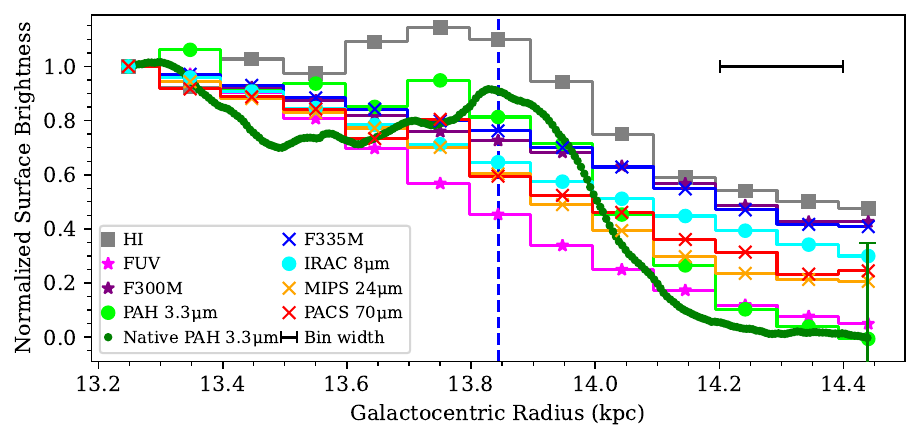}
\caption{Radial variations in brightness near the edge of M101's disk, normalized to the brightness of the innermost aperture (cyan in Figure~\ref{figure:prettypics}). The dashed blue line indicates the galactocentric radius where galaxy structure in the PAH \threethree\ map appears to end abruptly ($\sim$13.8~kpc). The propagated uncertainty in the \threethree\ map, including the uncertainty in the background subtraction, is indicated by the error bar on the rightmost point.}
\label{figure:radprofs}
\end{figure*}

\section{Radial Profiles}\label{sec:cliff}

One of the most striking visual features of the PAH \threethree\ maps of M101 is the `cliff' at the outskirts of the disk, where the galaxy structure traced by PAH \threethree\ seems to end abruptly (see Figure~\ref{figure:prettypics}). We investigated this quantitatively by comparing short radial profiles in this region of various photometric bands and emission line maps. In Figure~\ref{figure:radprofs}, we show radial emission profiles that range in wavelength from FUV to \textsc{Hi} 21\,cm extracted from maps at matched spatial resolution from the cyan regions shown in Figure~\ref{figure:prettypics}, which were arranged to avoid bright nearby H\textsc{ii} regions. We also dither these apertures by half their width to smooth out the profile.

The \textsc{Hi}, FUV, and PAH \threethree\ profiles are the only ones that show a significant change at the cliff radius; all others decrease smoothly through the cliff. At high metallicity, IRAC~8\um\ is dominated by emission from the 7.7\um\ PAH feature \citep{whitcomb23a}, but as the PAH bands weaken, an increasingly larger fraction of hot dust and starlight continuum contributes. PACS~70\um\ captures warm dust continuum and MIPS~24\um\ captures this and stochastically heated larger grain emission. Each shows a similar radial profile to IRAC~8\um, indicating that the IRAC~8\um\ trend is primarily tracing hot dust continuum and some starlight at these radii. This interpretation is further supported by the trends seen in F300M, F335M, and PAH \threethree. The radial profile of F335M follows approximately the same trend as the starlight-dominated F300M emission, and the cliff only becomes apparent when PAH \threethree\ is isolated by removing the underlying continuum in F335M.

The PAH \threethree\ profile at the native spatial resolution of F360M ($\sim3.9$\,pc) shows a similar but steeper trend than the convolved PAH \threethree\ profile (shown in dark green and light green in Figure~\ref{figure:radprofs}, respectively, though note the non-negligible error bar in the faint outskirts). This profile was calculated using 0$\farcs$2-wide apertures that are identical in azimuthal extent to the 7\arcsec-wide apertures shown in Figure~\ref{figure:prettypics}.

Taken together, these results imply that, rather than representing a threshold in metallicity, the PAH \threethree\ cliff is a result of both decreased PAH abundance and low overall ISM column density, similar to other diffuse areas across the disk, e.g. bubbles and inter-arm gaps. The cliff just happens to be the outermost, final arm at this azimuth in M101.

\end{appendix}

\bibliographystyle{aasjournal}
\bibliography{references}

\end{document}